\documentclass{emulateapj}
\def\stacksymbols #1#2#3#4{\def\theguybelow{#2}
        \def\verticalposition{\lower#3pt}
        \def\spacingwithinsymbol{\baselineskip0pt\lineskip#4pt}
        \mathrel{\mathpalette\intermediary#1}}
\def\intermediary #1#2{\verticalposition\vbox{\spacingwithinsymbol
        \everycr={}\tabskip0pt
        \halign{$\mathsurround0pt#1\hfil##\hfil$\crcr#2\crcr
                \theguybelow\crcr}}}
\def\lta{\stacksymbols{<}{\sim}{2.5}{.2}}
\def\gta{\stacksymbols{>}{\sim}{3}{.5}}

\begin{document}

\title{HEATING COOLING FLOWS WITH WEAK SHOCK WAVES}

\author{ William G. Mathews$^1$, 
Andreas Faltenbacher$^1$, \& Fabrizio Brighenti$^{1,2}$}

\affil{$^1$University of California Observatories/Lick Observatory,
Department of Astronomy and Astrophysics,
University of California, Santa Cruz, CA 95064\\
mathews@ucolick.org}

\affil{$^2$Dipartimento di Astronomia,
Universit\`a di Bologna,
via Ranzani 1,
Bologna 40127, Italy\\
brighenti@bo.astro.it}

\begin{abstract}
The discovery of extended, approximately spherical 
weak shock waves in the hot intercluster gas in 
Perseus and Virgo 
has precipitated the notion that these waves may be the 
primary heating process that explains why so little 
gas cools to low temperatures.
This type of heating has received additional 
support from recent gasdynamical models.
We show here that outward propagating, dissipating 
waves deposit most of their energy near the center of the 
cluster atmosphere. 
Consequently, if the gas is 
heated by (intermittent) weak shocks 
for several Gyrs, the gas within 30-50 kpc is heated 
to temperatures that far exceed observed values. 
This heating can be avoided if dissipating shocks are 
sufficiently infrequent or weak so as not to be the primary 
source of global heating. 
Local $PV$ and viscous heating associated with 
newly formed X-ray cavities are likely to be small, which is 
consistent with the low gas temperatures generally observed 
near the centers of groups and clusters where the 
cavities are located.
\end{abstract}


\section{Introduction}

Successful gasdynamical models of 
the hot virialized gas in galaxy groups and clusters must 
satisfy several cardinal requirements:

\begin{quote}
{\bf CR1} The rate that gas cools to low temperatures must be
less than $\sim 10$ percent of that predicted by conventional
cooling flows, ${\dot M} \sim L_x/(5 k T/2 \mu m_p)$, 
where $L_x$ is the bolometric X-ray luminosity and $T$ is the 
mean gas temperature.
\end{quote}

\begin{quote}
{\bf CR2} The gas temperature profile $T(r)$ 
must increase from the center 
out to $\sim$0.1-0.3 of the virial radius, 
i.e. $T(r)$ must look like a conventional cooling flow.
\end{quote}

\begin{quote}
{\bf CR3} The radial abundance profiles in the hot gas 
must agree with observation and 
in particular a mass $10^8 - 10^9$ $M_{\odot}$ 
of iron should be concentrated 
within $\sim$100 kpc of the central E or cD galaxy.
\end{quote}

\begin{quote}
{\bf CR4} 
It is necessary to identify 
the source of heating that reduces the 
baryon mass fraction below the cosmic value
in clusters/groups with mean temperatures less than about
3 keV.
\end{quote}

The first of these requirements is necessitated
by the well-documented 
absence of cooling gas in the X-ray spectra of
galaxy groups and clusters 
(e.g. Peterson et al. 2001; B\"ohringer et al. 2002).
Evidently some additional source of heating is required.
In view of the limited energy available in supernova explosions 
and the uncertain efficiency of thermal conduction 
(Voigt \& Fabian 2004), 
it is generally thought that the cluster 
gas is heated directly or indirectly
by jets or energetic outflows originating near 
supermassive black holes (AGN) in the cores of cluster-centered
elliptical galaxies.
But there is no general consensus regarding the physical processes
responsible for the heating or the means by which
energy is delivered to gas at large distances from the central AGN.
In any case, the X-ray emitting gas closest to the central AGN must 
not be strongly heated since that would violate {\bf CR2}. 
Positive temperature gradients within $0.1 - 0.3 r_{vir}$ 
are typical among cooling flow clusters and groups
(Allen, Schmidt \& Fabian 2001; De Grandi \& Molendi 2002),
although the temperature profile in a few clusters is 
more nearly isothermal (e.g. AWM-4; O'Sullivan et al.  2005). 
Assuming that the hot cluster gas is in approximate hydrostatic 
equilibrium in an NFW mass distribution, 
gasdynamical models that produce correct temperature profiles 
will also have correct density and entropy profiles.

Except possibly for the very center, the gas phase iron abundance 
in groups and clusters generally decreases from $\sim$solar 
to $\sim0.3 \pm 0.1$ solar at the outer limit of observations. 
In cool-core clusters a significant fraction of the total 
gas phase iron mass, $M_{Fe} \approx 10^8 - 10^9 M_{\odot}$, is 
concentrated within $\sim100$ kpc of the 
central E or cD galaxy (de Grandi et al. 2004).
Metal abundance gradients in the hot gas retain an integrated record 
of its past association with stars and supernovae that must 
be reproduced by any successful model ({\bf CR3}).

To explore the gasdynamical consequences of heating,
Brighenti \& Mathews (2002a; 2003) considered 
a wide variety of {\it ad hoc} heating scenarios and described the
effect of each on the hot gas over many Gyrs. 
In these models the gas was heated 
either continuously or intermittently, 
either symmetrically near the center or at locations off center. 
X-ray cavities formed in the heated gas 
generated weak shock waves similar to those found in Perseus. 
We concluded at that time that none of the many types of heating 
scenarios satisfied all cardinal requirements 
listed above. 
For example, if the heating is symmetric around the center of the flow 
at a level sufficient to quench 
the radiative cooling there, we found that cooling still 
occurred at larger radii in the flow (violating {\bf CR1}) or 
the radial temperature gradients in the heated region were invariably 
negative (violating {\bf CR2}). 
To fully shut down a cooling flow it is necessary to heat the 
gas nearly to the (``cooling'') radius $r_{cool}$, the 
radius at which the gas would cool in $\sim 10^{10}$ years 
or during the cluster lifetime.

One of the principal questions in understanding the 
long term evolution of cluster gas is whether mass as well as energy 
must be transported out from the central AGN. 
To investigate this possibility, 
we recently considered idealized ``circulation flows'' in which 
gas heated near the center is buoyantly transported to large radii
in the flow where it merges with the ambient gas 
and subsequently flows back in (Mathews et al. 2003; 2004).
In these circulation flows gas moves 
simultaneously in both radial directions -- 
there is no net flow of matter and no radiative cooling 
to low temperatures (satisfying {\bf CR1}). 
The dense inflowing gas emits most of the X-ray emission and 
resembles a traditional cooling flow that fills only a fraction of the 
available volume.
The remaining volume is occupied by outflowing 
low-density bubbles of heated 
gas that do not contribute much to the X-ray spectrum. 
A single-temperature interpretation of the emission from such 
a flow is dominated by the inflowing gas and resembles 
a normal cooling flow (satisfying {\bf CR2}). 
Furthermore, we showed that 
successful circulation flows must carry both mass and energy 
out from the center to approximately the cooling radius. 
The iron abundance peak that surrounds the central E galaxy 
is easily explained with circulation flows as the repository 
of all the iron produced by Type Ia supernovae in the central 
galaxy over time (satisfying {\bf CR3}). 

Other heating scenarios can also be imagined. 
In the discussion below we explore the possibility that 
weak dissipating waves created near the center of the flow 
propagate throughout the hot gas, 
globally heating the gas and 
quenching the radiative cooling to low temperatures.

The deep 200ks {\it Chandra} observation of the Perseus Cluster 
by Fabian et al. (2003) revealed the presence of 
an ensemble of approximately spherical weak 
shock waves (``ripples'') within about 50 kpc of the center. 
The Perseus cluster also contains at least four 
X-ray cavities (bubbles) all within about 35 kpc of the 
center of the cluster (Birzan et al. 2004).
Forman et al. (2003) describe similar spherical waves at
14, 17 and 37 kpc in the Virgo cluster. 
These distant shock-ripples are almost certainly created by 
the slow, intermittent inflation of new bubbles near the 
central AGN. 
It is important to recognize that 
weak shocks are produced even by bubbles that expand subsonically. 
The ripples observed have progressed about half way to the cooling 
radius which is about 130 kpc in Perseus 
and 75 kpc in Virgo (Peres et al. 1998),
assuming $H_0 = 70$ km s$^{-1}$ Mpc$^{-1}$. 

Fabian et al. (2003) argue that dissipating shocks  
(ripples) excited by inner bubbles are likely to 
be the dominant heating mechanism that balances 
radiative cooling throughout the hot gas. 
They claim that alternative modes of heating -- large 
numbers of tiny effervescent bubbles (Begelman 2001), 
major central eruptions (Soker et al. 2001; Kaiser \& Binney 
2002) or heating near the buoyant bubbles 
(Churazov et al. 2001, 2002; 
Quilis et al. 2001; 
Br\"uggen et al. 2002; Br\"uggen \& Kaiser 2002;
Br\"uggen 2003) -- are less effective within $r_{cool}$ 
than the quasi-continuous large scale dissipative heating by 
shock-ripples, which can persist even in those clusters 
(e.g. A1835) that do not currently have central bubble activity. 

The idea that ripple wave heating is dominant has been 
amplified recently by Clarke et al. (2005) who claim that 
``these shocks dissipate energy which may be sufficient to 
balance the effects of radiative cooling in the cluster cores.''
This conclusion has been strongly advocated by 
Ruszkowski, Br\"uggen \& Begelman (2004a,b) 
and Dalla Vecchia et al (2004), who argue 
that the approximately spherical dissipation of 
large-scale waves is more efficient in globally balancing 
the radiative losses in cluster gas within $r_{cool}$ 
than viscous heating associated with the flow 
just around the X-ray cavities. 
They also emphasize the importance of 
AGN intermittency in exciting individual ripple waves. 

Even with the best available observations, it is impossible to detect
an increase in the gas temperature or thermal energy across
individual ripples. The two ripples observed in Perseus, for example,
are insufficient to increase the cluster energy (or entropy) enough to
shut down the global cooling for many Gyrs.
Because the amount of energy carried and entropy dissipated by
each wave is so small, it is necessary to perform a calculation such
as we describe here
to test the idea that a large number of ripples sustained over
time can provide enough dissipation to stop the cooling -- without
destroying the temperature profile as observed. We show that 
wave-heating fails because the central regions of the cluster, 
through which all outgoing waves must pass, receive too much 
dissipated energy and after a few Gyrs become much hotter 
than the temperature observed.

It is not surprising that
Dalla Vecchia et al (2004) and
Ruszkowski, Br\"uggen \& Begelman (2004a,b)
were able to reduce radiative cooling by ripple heating
in their 2D and 3D simulations. 
To achieve this it is only necessary to adjust 
the (highly uncertain) power delivered to 
the hot gas by the AGN to be comparable to 
the bolometric X-ray luminosity $L_x$. 
By this {\it ad hoc} means radiative cooling can be shut down 
for a time $\sim t_{cool}(r_{heat})$ within the radius 
$r_{heat} \le r_{cool}$ where the AGN energy is deposited 
by ripple shock waves.
However, it is essential to determine whether or not 
cooling flows can be suppressed by ripple-heating 
over many Gyrs 
and still satisfy the gas temperature profiles 
generally observed ({\bf CR2}). 
This is the question we address here. 
To establish with certainty that 
ripple wave dissipation can balance radiative losses 
in a cooling flow, 
it is necessary to calculate for 
times comparable to the relevant cooling time or evolutionary 
lifetime of the cluster. 

\subsection{Viscous Heating by Shocks, Sound Waves 
and Turbulence}

There has been some confusion 
about the magnitude of the viscosity that is required 
for viscous dissipation to offset radiative losses
in the hot gas in groups and clusters.
In the absence of magnetic fields, the viscosity  
$\mu_s = 1.1 \times 10^{-16} T^{5/2}$ g cm$^{-1}$ s$^{-1}$
depends only on the temperature (Braginskii 1958). 
Viscous heating depends strongly on the local 
velocity gradient, $\sim\mu (d u/d\ell)^2$
erg cm$^{-3}$ s$^{-3}$ 
where $\ell$ is a spatial coordinate 
appropriate to the scale of the region where 
dissipation occurs.
Although viscous dissipation occurs on small 
spatial scales comparable to the mean free path, 
it can heat a large volume of gas 
by the passage of dissipating wave fronts or by 
widespread turbulence.

X-ray cavities can heat the ambient gas 
by performing work and creating waves as they 
are initially formed and by viscous dissipation of turbulence 
produced by their buoyant motion.
Birzan et al. (2004) estimated the $PV$ work done by
expanding cavities on the hot gas in 16 galaxy
clusters and concluded that ``this mechanism
alone probably does not provide
a general solution to the cooling problem.''
If X-ray cavities inflate adiabatically into an
invisid gas, an amount of work $PV$ is
transferred to the local gas.
However, a significant fraction of this work is
expended in displacing local gas outward in the
cluster potential, but this gas returns to its original
radius after the buoyant bubble moves away, performing
a nearly equal amount of negative work, $-PV$.
Consequently $PV$ overestimates the total
energy received by the ambient gas from the cavities.

The absence of resonance scattering of X-ray lines
led Mathews et al. (2001) and 
Churazov et al. (2004) to suggest
considerable subsonic turbulence in the central
$\sim 60$ kpc of Perseus
but strong, extended turbulence may be inconsistent
with steep iron abundance gradients observed in
Perseus and many similar clusters
(e.g. Fabian et al. 2003).
If such turbulent energy exists, the buoyant motions 
of X-ray cavities are a likely source. 
It is interesting to compare the rate of 
viscous dissipation near a cavity moving 
at $\sim 30$ kpc in Perseus ($T \approx 3 \times 10^7$ K;
$n_e \approx 0.03$ cm$^{-3}$)
to the local volume radiative emissivity.
The drag force on a bubble (cavity) of radius $r_b$
moving at velocity $u_b$ is $F_d = \delta_b \rho u_b^2 \pi r_b^2$
where $\delta_b < 1$ is the drag coefficient.
If the power expended against drag by a single bubble is $F_d u_b$,
the total power within radius $r$ in the flow is 
$\delta_b \rho u_b^3 \pi r^2 f_a^2$ where 
$f_a = N_b \pi r_b^2/4 \pi r^2 < 1$ is 
the area filling factor for $N_b$ 
bubbles at radius $r$ in the flow.
The total power deposited in volume $4 \pi r^3 /3$, 
$(3/4) \delta_b \rho u_b^3 f_a^2/r$, is ultimately 
dissipated by viscosity at a rate 
\begin{displaymath}
{\dot \varepsilon}_{visc} =
\delta_b {\cal M}^3 f_a^2 {3 \over 4} {\rho c_s^3 \over r}
= 2.6 \times 10^{-25} \delta_b {\cal M}^3 f_a^2
\end{displaymath}
\begin{displaymath}
\times \left( {T \over 3 \times 10^7~{\rm K} } \right)^{3/2}
\left({ n_e \over 0.03~{\rm cm}^{-3}}\right)
\left( { r \over 30~{\rm kpc}} \right)^{-1}
~~~{{\rm erg} \over {\rm cm}^3~{\rm s}}
\end{displaymath}
where we assume that the bubbles move subsonically, 
i.e., ${\cal M} = u_b/c_s < 1$
where $c_s = (\gamma k T /0.61 m_p)^{1/2}$ is the
sound speed.
Note that the absolute value of the viscosity
determines the small scales of energy dissipation but
not the amount of turbulent energy dissipated.
The rate that energy is radiated at this radius is 
\begin{displaymath}
{\dot \varepsilon}_{rad} =
\left({ \rho \over m_p}\right)^2 \Lambda(T,z)
\approx 2.4 \times 10^{-26} 
\end{displaymath}
\begin{displaymath}
\times \left({ n_e \over 0.03~{\rm cm}^{-3}}\right)^2
\left({\Lambda \over 2 \times 10^{-23}}\right)
~~~{{\rm erg} \over {\rm cm}^3~{\rm s}}.
\end{displaymath}
Since we expect $\delta_b {\cal M}^3 f_a^2 \ll 1$, the turbulent 
viscous dissipation directly associated with bubble buoyancy 
is unlikely to equal the local radiative losses. 

Viscous dissipation $\mu (d u/d\ell)^2$ and 
heating can also be important in 
large amplitude sound waves and weak shock waves 
and must be investigated as a possible means of heating
cluster gas.
When ripple-waves were discovered in Perseus and Virgo,
it was assumed that they were weak shocks of the sort 
that accompanies the creation of X-ray cavities 
(e.g. Brighenti \& Mathews 2002b).
Recently Fabian et al. (2005) propose that cluster gas 
can be heated by linear sound waves of frequency 
$f \sim (3 \times 10^5~{\rm yr})^{-1} $ that dissipate on scales  
$$
\ell_{diss} = {2 \rho c_s^3 \over (4\mu /3)}{1 \over (2\pi f)^2}
\sim 50~{\rm kpc}.
$$
Fabian et al. show that it is possible to approximately offset 
radiative cooling in large clusters if the wave 
frequency is sufficiently high and the sound wave luminosity 
at this frequency is sufficiently large. 
The low amplitude 
linear waves required for this type of heating 
probably cannot be detected by X-ray observations. 
By comparison, in shock waves the dissipation occurs 
over a few mean free paths regardless of the 
magnitude of the viscosity, 
i.e. unlike linear wave dissipation 
the shock jump conditions and entropy increase do not 
explicitly depend on the viscosity.

An important question that has not been throughly considered 
is the competition between the 
viscous dissipation of large amplitude 
sound waves as they decay with no change in the wave form 
and the progressive steepening of the wave profile that 
results in a shock wave.
The dimensionless ratio that determines 
these quite different evolutionary changes in the 
wave form is the Goldberg number 
$$
\Gamma = {(\gamma + 1){\cal M}/2 \over \ell_{diss}^{-1}/k}
$$
where $\delta u = {\cal M} c_s$ is the velocity amplitude 
of the wave and $k = 2 \pi /\lambda$
(e.g. Kinsler et al. 2000). 
The dimensionless Goldberg number is the ratio of 
${\cal M}$, an indicator of the degree of non-linearity 
in the wave, to $\ell_{diss}^{-1}/k$,
the amplitude attenuation over one wavelength 
at the fundamental frequency of the wave form.
The evolution of waves created by X-ray 
cavities in the Perseus cluster with 
period $T_p$ can be estimated by 
combining the various expressions above,
$$
\Gamma = 350 {{\cal M} \over f_{\mu}} 
\left( { n_e \over 0.03{\rm cm}^{-3} } \right)
\left( { T_p \over 3 \times 10^7{\rm yr} } \right)^3
$$
$$
\times \left( { 3 \times 10^7{\rm K} \over T } \right)
$$
where $f_{\mu} = \mu/\mu_s \lta 0.2$ is the 
deviation of the viscosity from the Braginskii-Spitzer value.
Since ${\cal M}$ is less than (but comparable to) unity 
for subsonically expanding cavities, it is obvious 
that $\Gamma \gg 1$ so the waves created by expanding  
X-ray bubbles must steepen rapidly into weak shocks.

\subsection{Density Profile and Shock Wave Dissipation}

We describe below how hot cluster gas is heated 
by dissipating waves created near the central AGN. 
Since the gas density in groups and clusters 
varies less steeply than 
$\rho \propto r^{-2}$, expanding ripple 
waves created in the bubble 
region ($\sim10 - 20$ kpc) lose amplitude 
as they encounter an increasing mass per 
unit area; this loss in wave amplitude occurs 
in addition to viscous dissipation. 
We wish to determine 
if the temperature profiles typically observed in clusters 
are consistent with ripple-shock heating to 
radius $r_{heat}$ where $t_{cool}(r_{heat})$ is 
comparable to the likely age of the cluster.
Specifically, we study wave dissipation in the Perseus cluster, 
where weak ripple-shocks are well-documented by observation. 
We begin with a discussion of the properties 
of this cluster.

\section{The Perseus Cluster}

Churazov et al. (2004) give analytic fits 
to the density and temperature profiles in Perseus 
at the spatial resolution of their 
{\it XMM-Newton} observations. 
The deep {\it Chandra} observation of of Perseus by 
Sanders et al. (2004) is consistent with the 
same temperature profile as Churazov et al.,
\begin{equation}
T = 8.12 \times 10^7 
{ [1 + (r/71)^3] \over [2.3 + (r/71)^3]} ~ K 
\end{equation}
where $r$ is in kpc with a distance determined
with $H_0 = 70$ km s$^{-1}$ Mpc$^{-1}$.
The azimuthally averaged 
density profile in Perseus is given by 
\begin{displaymath}
n_e(r) = {0.0192 \over [1 + (r/18)^3]}
\end{displaymath}
\begin{equation}
~~~~~~+ {0.046 \over [1 + (r/57)^2]^{1.8}}
+ {0.0048 \over [1 + (r/200)^2]^{0.87}}
~{\rm cm}^{-3}.
\end{equation}
The last two terms are taken from 
Churazov et al. (2004) and the 
first term corrects for the {\it Chandra}
observations of Sanders et al. in the 
high density core.

The acceleration of gravity $g(r)$ in 
the Perseus cluster 
can be found by assuming hydrostatic 
equilibrium in the hot gas 
and differentiating the equations above,
\begin{equation}
g_{nT}(r) = - {k T \over \mu m_p}
\left( {1 \over n_e} {d n_e \over dr}
+ {1 \over T} { dT \over dr}\right),
\end{equation}
and this is shown with the dot-dashed line in Figure 1.
Within about 10 kpc 
the observed acceleration $g_{nT}(r)$ decreases 
since $T(r)$ and $n_e(r)$ in Equations (1) and (2) 
become constant as $r \rightarrow 0$. 
A centrally decreasing gravitational acceleration 
is clearly inconsistent with the expected dark 
and luminous mass distributions in this region. 
The azimuthally (i.e. spherically) 
averaged fitting functions for 
$n_e(r)$ and $T(r)$ fail for $r \lta 10$ kpc 
since this is the region occupied by large 
X-ray cavities. 
Even if the gas is close to hydrostatic 
equilibrium everywhere, as we believe to be 
the case, azimuthal averaging underestimates 
$n_e$ and $d\log n_e /dr$ in 
the high density intercavity gas that produces 
most of the X-ray emission. 
Although the drop in $g_{nT}(r)$ within 10 kpc is 
unphysical, in the region $10 \lta r \lta 300$ kpc, 
where both $n_e(r)$ and $T(r)$ are observed, 
$g_{nT}$ should accurately measure the Perseus 
gravitational field.

We match the observed $g_{nT}(r)$ 
by combining an NFW dark 
halo and a stellar contribution 
from the cluster-centered galaxy NGC 1275, 
$g = g_{nfw} + g_*$.
The halo acceleration is simply
\begin{equation}
g_{nfw} = {G M_{vir} \over r^2} 
{\log(1 + y) - y/(1+y) \over \log(1+c) - c/(1+c)}
\end{equation}
where $y = cr/r_{vir}$, $c$ is the concentration,
and $r$ is the radius in kpc. 
The acceleration $g_{nfw}(r)$, determined with parameters 
$M_{vir} = 8.5\times 10^{14}$ $M_{\odot}$,
$r_{vir} = 2.440$ Mpc, and 
$c = 6.81$, and shown as a long-dashed line in 
Figure 1 is a reasonably good fit to $g_{nT}(r)$ 
at $10 \lta r \lta 300$ kpc.

The physical properties of the central galaxy 
in Perseus, NGC 1275, are problematical since 
it appears to be an elliptical galaxy that has 
experienced a recent merger, perhaps with a 
smaller spiral galaxy 
(Conselice et al. 2001).
The global color of NGC 1275, 
$(B - V)_o = 0.72 \pm 0.03$, 
is significantly bluer than a normal E galaxy, 
but the stellar surface brightness profile 
follows a de Vaucouleurs $R^{1/4}$ law to 
at least $150$ kpc 
(Prestwich et al. 1997).
We therefore assume here that the stellar 
mass of NGC 1275 is 
dominated by an old stellar population, as in 
normal E galaxies, intermixed with an additional 
population of young, luminous stars that 
does not contribute significantly to the total mass. 
The stellar mass distribution $M_*(r)$ and $g_*(r)$ 
for a de Vaucouleurs profile
can be found from the effective radius,  
the total luminosity $L_B$ of {\it old} stars, 
and the stellar mass to light ratio $M/L_B$.
At a distance of 73.4 Mpc 
the effective radius of NGC 1275 is 
$R_e = 6.41$ kpc (RC3). 
The total magnitude $M_V = -21.62$ 
of the old stars can be found 
from the stellar velocity dispersion 
$\log \sigma_* = 2.40$ (Bettoni et al. 2003)
using the $\log \sigma_* - M_V$ correlation from 
Faber et al. (1997).
If $B - V = 0.95$, as in normal E galaxies 
with old stellar populations, then 
$M_B = -20.67$ and $L_B = 2.70 \times 10^{10} L_{B\odot}$, 
and we ignore the mass and luminosity of recently formed stars. 
For this $L_B$ we find 
$M/L_B =  9.0$ from 
Trujillo, Burkert \& Bell (2004) 
so the total stellar mass of NGC 1275 is 
$M_{*t} = 2.43 \times 10^{11}$ $M_{\odot}$.
The stellar acceleration $g_*(r) = G M_*(r)/r^2$ 
for a de Vaucouleurs profile, shown as the 
short-dashed line in Figure 1, can be accurately fit with 
\begin{displaymath}
g_*(r) = 
\left[ \left( {r^{0.5975} \over 3.206 \times 10^{-7}}\right)^s
+ \left( {r^{1.849} \over 1.861\times 10^{-6}}\right)^s 
\right]^{-1/s}
\end{displaymath}
in cgs with $s = 0.9$ and $r$ in kpc. 
We neglect the self gravity of the hot gas. 
The total two-component acceleration 
$g(r) = g_{nfw} + g_*$, shown as a solid line 
in Figure 1, fits $g_{nT}(r)$ quite well for 
$r \gta 10$ kpc. 

\section{Exciting Waves}

We wish to generate spherical shock waves of 
low amplitude that propagate out through the Perseus gas. 
The objective is to determine if this type of 
shock heating (dissipation) can provide enough energy 
to the cluster gas to shut down the cooling for times 
$\sim4$ Gyrs comparable to the likely age of the cluster and 
still preserve the observed temperature profile.
Waves can be generated with a sinusoidally 
oscillating piston located in the bubble region. 
To create such a AGN wave machine 
it is convenient to use Lagrangian rather than 
the usual Eulerian hydrodynamics since the boundary 
condition at the piston is more straightforward. 
The equation of motion is 
\begin{equation}
{\partial u \over \partial t} 
= - {1 \over \rho} {\partial \over \partial r}(P + Q) - g
\end{equation}
where
\begin{equation}
u = {\partial r \over \partial t},
\end{equation}
is the velocity defined in terms of the local 
Eulerian radius $r$.
The equation of continuity is simply 
\begin{equation}
\rho \Delta (4 \pi r^3/3) = \Delta m
\end{equation} 
where $\Delta (4 \pi r^3/3)$ is the volume of 
computational zones of mass $\Delta m$. 
The thermal energy equation with radiative losses is 
\begin{equation}
{\partial \varepsilon \over \partial t}
={P + Q \over \rho^2} {\partial \rho \over \partial t}
- \left({\rho \over m_p}\right)^2 \Lambda(T,z)
\end{equation}
where $\varepsilon = 3P/2\rho$ is the 
specific thermal energy and $z \approx 0.4$ is the metal abundance 
in solar units.
The artificial viscosity 
\begin{equation}
Q = a^2 \rho (\Delta u)^2
\end{equation}
depends on the 
velocity difference $\Delta u$ across the computational 
zones and a dimensionless coefficient $a$ 
of order unity to smooth the post-shock flow. 
The artificial viscosity ensures that the correct dissipative 
heating in the shock occurs over several computational 
zones.

Spherical ripple waves can be generated by requiring 
that the inner boundary $r_p(t)$ of the innermost computational 
zone, initially at $r_{p0}$, oscillates 
with amplitude $\Delta r_p$ and period $T_p$,
\begin{equation}
r_p = r_{p0} + \Delta r_p \sin(2 \pi t/T_p)
\end{equation}
with velocity
\begin{displaymath}
u_p = (2 \pi \Delta r_p/T_p)\cos(2 \pi t/T_p).
\end{displaymath}
Gas within the initial piston radius $r_{p0}$ does not 
participate in the flow -- this is reasonable 
since we are interested in heating the much larger volume 
of gas between the 10-30 kpc region where the cavities 
are observed and the distant cooling radius, 
$r_{cool} \approx 130$ kpc.
The total mechanical energy delivered to the gas 
by the piston after time $t$ is
\begin{equation}
E_m(t) = \int_0^t 4 \pi r_p^2 P_p u_p dt
\end{equation}
where $P_p(t)$ is the instantaneous gas pressure at $r_p(t)$.
If $P_p$ were constant then no mechanical luminosity 
(power) is delivered to the gas in one cycle,
\begin{displaymath}
\langle L_m \rangle_{cy} = \langle dE_m/dt \rangle_{cy} =
\end{displaymath}
\begin{displaymath}
{8 \pi^2 r_{p0}^2 \Delta r_p P_p \over T_p} 
\int_0^{2\pi} [1 + (\Delta r_p/r_{p0})\sin \tau]^2 
\cos \tau d\tau = 0.
\end{displaymath}
The mechanical luminosity depends critically on the 
asymmetry in the gas pressure 
at the piston $P_p(t)$ which is significantly larger
when the piston moves outward, compressing the 
local gas, than when it recedes inward. 
Nevertheless in general we expect 
$L_m \propto r_{p0}^2 \Delta r_p / T_p 
\propto r_{p0}^2 |u_p|$. 
The piston parameters -- $r_{p0}$, 
$\Delta r_p$, and $T_p$ -- can be selected by 
choosing $T_p \sim 10^7 - 10^8$ years, 
as expected for the AGN periodicity from many observations, 
and then selecting $\Delta r_p$ so that the peak piston velocity 
$|u_p| = 2 \pi \Delta r_p / T_p$ remains subsonic 
since strong shocks are not observed around the X-ray cavities.
Finally $r_{p0}$ can be increased until $\langle L_m \rangle$ 
is comparable to the bolometric luminosity $L_x$, 
a necessary condition to quench the cooling flow. 
Alternatively, we can compare $E_m(t)$ with the total 
energy radiated after time $t$,
\begin{displaymath}
E_{rad}(t) = \int_0^t dt \int_V (\rho/m_p)^2 \Lambda(T,z) dV,
\end{displaymath}
where $V$ is the total computational volume.

\section{A Pure Cooling Flow in Perseus}

Before considering the heating effects of outwardly 
propagating waves, it is useful to calculate a pure 
cooling flow in the Perseus cluster environment.  
This can be accomplished by fixing the piston at 
some small radius $r_{p0} = 1$ kpc and setting $\Delta r_p = 0$. 
In our solutions of the Lagrangian equations we begin 
with a configuration that is in strict hydrostatic 
equilibrium, which is probably an excellent approximation 
outside the bubble region and a good approximation even near 
the center where the gas velocities are largely subsonic.
Initial hydrostatic equilibrium is enforced on the 
computational grid by demanding that the differenced 
form of equation (5) corresponds to zero velocity 
$u_j = 0$ at all zone boundaries $r_j$,
\begin{displaymath}
g_j = - {1 \over 0.5(\rho_{j + 1/2} + \rho_{j - 1/2})}
\end{displaymath}
\begin{equation}
\times {k \over \mu m_p}
{(\rho_{j + 1/2} T_{j+1/2} - \rho_{j - 1/2}T_{j-1/2})
\over 0.5(r_{j+1} - r_{j-1})}.
\end{equation}
The gravitational acceleration at 
every zone boundary is $g_j = g(r_j)$, 
where $g(r)$ is plotted in Figure 1 with a solid line. 
The observed temperature profile is insensitive to 
the volume filling factor in the bubble region 
(Mathews, Brighenti \& Buote 2004)
and {\it Chandra} and {\it XMM} observations are 
consistent with the same $T(r)$, so it is natural 
to assume $T_{j+1/2} = T(r_{j+1/2})$, using the 
analytical fit to the observed temperature profile 
described previously.
With these assumptions, the difference equation 
above can be solved recursively for $\rho_{j+1/2}$ 
from the innermost zone outward. 
If necessary, the density of the innermost zone 
is adjusted slightly so that the overall density profile 
agrees with that described by equation (2).
In general the initial gas density $\rho_{j+1/2}$ 
within $r \sim 10$ kpc, which corresponds to the 
intercavity density, 
is somewhat larger than the analytic fit (Eqn. 2), which 
is a spherical average including both cavity 
and intercavity gas. 

As gas radiates and flows toward the center 
in a normal cooling flow, the density 
rises until radiative cooling eventually becomes catastrophic.
When the gas temperature in a Lagrangian zone 
drops below $3 \times 10^5$ K, 
the zone becomes spatially very narrow and 
we remove it from the calculation, filling its small 
volume with a small amount of gas from neighboring zones. 
Since the inner boundary at $r_{p0} = 1$ kpc 
is stationary, all the gas cools just beyond that small radius.  
A shock near $r_{p0}$ prevents the inflowing gas from 
attaining large negative velocities.

The solid lines in Figure 2 show the cooling flow after 
4 Gyrs compared to the initial profiles shown as 
dashed lines. 
We select 4 Gyrs (redshift $z \sim 0.5$) since massive 
clusters such as Perseus may have evolved significantly 
over longer times.
Toward the center of the flow the computed 
gas density $n_e$ rises 
above the observations in $r \lta 20$ kpc
as $r \rightarrow r_{p0}$; 
such a density rise is typical in traditional 
cooling flows (see e.g. Mathews \& Brighenti 2003) 
and is enhanced further 
in Perseus by the central X-ray cavities 
that lower the mean observed density.
By 4 Gyrs the gas cooling rate shown in 
Figure 3 has stabilized 
at ${\dot M} = 250$ $M_{\odot}$ yr$^{-1}$ and the huge total 
mass of cooled gas at the center,
$M_{cool} = 9.8 \times 10^{11}$ $M_{\odot}$, is equal to 
all the gas within 73 kpc in the initial configuration.
Our ${\dot M}$ is nearly identical to the cooling flow rate 
for Perseus predicted by Sanders et al. (2004),
${\dot M} \approx 255$ $M_{\odot}$ yr$^{-1}$, 
but is inconsistent with their {\it Chandra} observations
that show no evidence that gas is 
cooling below $\sim 2 \times 10^7$ K. 
Heating by ripple wave dissipation or other means is 
necessary to prevent this cooling.

We have advocated that gas near the centers of 
cooling flows is heated and buoyantly recirculated to 
large radii extending to $\sim r_{cool}$
(Mathews et al. 2003; Mathews, Brighenti \& 
Buote 2004).
In these ``circulation flows'' both mass and energy 
are transported outward, but the thermal profile of 
of the flow is nearly identical to that of a normal cooling 
flow, i.e. $T(r)$ 
is insensitive to the radial variation of the 
volume filling factor 
occupied by the rising cavities (bubbles) 
as shown by Mathews et al. (2004).
For this reason it is noteworthy in Figure 2 that 
the computed  cooling flow 
temperature profile after 4 Gyrs is almost identical 
to the observed profile for $r \gta 30$ kpc. 
Since a pure cooling inflow matches the observed temperature  
and density in this outer region of Perseus so well,
we infer that there is no need for conductive heat 
transport in this region of the flow, although some modest 
heating may occur. 
In the following section we explore whether this 
radiative cooling in the Perseus cluster,
${\dot M} = 250$ $M_{\odot}$ yr$^{-1}$,
can be significantly reduced 
by dissipative losses in outwardly propagating 
weak shock waves (ripples).

\section{Ripple-Heated Flows}

To create a series of
outward propagating weak shock waves that move out
into the Perseus atmosphere,
we allow the spherical piston at radius $r_{p0}$ to 
undergo sinusoidal oscillations with amplitude 
$\Delta r_p$ and period $T_p$.
Since strong shocks are rarely observed surrounding
X-ray cavities, 
we select parameters for which the maximum velocity 
of the piston $|u_{p,max}| = 2 \pi \Delta r_p/T_p$
is less than the sound speed in the local hot gas,
$c_s(r_p) = [5 k T(r_p)/3\mu m_p]^{1/2} 
= 958 [T(r_p) / 3\times 10^7~{\rm K}]^{1/2}$ 
km s$^{-1}$. 
Note that weak shocks are created even when $|u_p| < c_s$.  

The flows with piston parameters listed in Table 1 are 
a representative sample of our exploration of heating 
by dissipating ripple waves.
The first flow of this type is shown 
in Figure 4 at two times, 
designated $a1$ after $4\times 10^7$ yrs and $a$ after 4 Gyrs. 
If the gas density varied as $\rho \propto r^{-2}$, 
outwardly propagating dissipationless 
waves would encounter nearly the same mass 
per unit area at every radius and their amplitude would 
remain unchanged.
However, in Perseus the density drops slower than 
$\rho \propto r^{-1.6}$ so the wave amplitude is expected
to decrease with radius as a larger mass participates in the waves. 
This is seen most clearly in the gas velocity 
panel in Figure 4 for model $a$. 
The piston parameters of model $a$ have been selected so that 
the mechanical energy $E_m(t)$ expended by time $t$ 
is initially slightly larger than 
the radiated energy, as illustrated in Figure 5 -- this 
requires a piston velocity $|u_p|$ that is a large 
fraction of the initial sound speed.
As can be seen in Figure 3,
the total mass of cooled gas in model $a$ is small
and this cooling occurs close to the piston.
Nevertheless, this solution is clearly unacceptable
because it has the wrong density and temperature profiles.
As expected, most of the wave dissipation occurs 
at small radii in the flow, resulting in temperatures 
exceeding $10^8$ K with negative $dT/dr$ continuing to 
nearly the cooling radius. 
The view of model $a$ at an earlier time in the 
first column of Figure 4 (model $a1$), 
shows that rather strong shocks 
were present at early times. 
By 4 Gyrs these shocks 
produced the heating shown in the second column 
(model $a$).
After the central regions in Perseus have been heated 
to these unrealistically high temperatures, the 
piston produces a 
more rounded innermost wave profile 
in flow $a$, a subsonic non-linear wave that shocks 
further out.

In model $b$ we consider a flow with the same velocity $|u_p|$ 
as in model $a$ but with a shorter period $T_p = 10^7$ yrs 
and the resulting flow has an entirely different character.
The outer part of flow $b$ beyond 50 kpc 
in Figure 4 is essentially a 
cooling inflow, and this is verified in the bottom 
panel where $\langle u \rangle < 0$. 
The cooling rate and total cooled gas in model $b$ 
are unacceptably large (Figure 3).
All of this cooling 
occurs at $r \approx 50$ kpc where the gas temperature has 
a sharp minimum and the density has a corresponding rise. 
The waves generated by the piston are unable to penetrate 
the impedance barrier at $r \approx 50$ kpc and this accounts 
for the very low wave amplitude at larger $r$.
Within $r \approx 50$ kpc the dissipating waves reflect back and 
forth raising the gas temperature far above that observed 
in Perseus.

In models $c - f$ the piston amplitude 
$\Delta r_p$ and period $T_p$ are chosen so that 
the piston velocity $|u_p|$ is more subsonic than the 
previous solutions (Table 1).
As we have discussed already,
to achieve an approximate balance between mechanical and 
radiative luminosity when $|u_p|$ is smaller, it it necessary
to increase the mean piston position $r_{p0}$. 
In model $c$ the piston position 
$r_{p0} = 31.5$ kpc is near the outer limits 
of the observed X-ray cavities in Perseus.
In Figure 5 it is seen that $E_m$ is only slightly 
less than $E_{rad}$ during the entire calculation to 4 Gyrs, 
but Figure 3 shows that very little gas 
(all near the piston) has cooled.
However, as with model $a$ the gas temperature 
in the region $37 \lta r \lta 57$ kpc 
of model $c$ exceeds $10^8$ K, 
which is far in excess of the temperature 
observed in Perseus. 

Finally, we consider a series of models, $d - f$ in which 
$|u_p|$ remains fixed at the small value in model $c$, 
but $r_{p0}$ is slowly reduced. 
A smaller mean piston radius $r_{p0}$ 
is expected to result in lower 
mechanical energy input $E_m \propto r_{p0}^2|u_p|$ relative 
to the total radiated energy $E_{rad}$.
This increasing discrepancy 
can be seen by comparing models $c$ and $f$ in Figure 5.
The overall character of solutions $c \rightarrow e$ remains 
about the same, with a small amount of cooling 
(Figure 3) but with temperature profiles that peak 
at small radii and therefore 
are inconsistent with the observations.
Finally, in model $f$ the mechanical energy is 
unable to keep large amounts of gas from cooling (Figure 3). 
Unlike the flows $c - e$, most of the cooling in flow $f$ 
occurs not at the piston but in the cooling cusp 
at $r \approx 30$ kpc. 
Model $f$ resembles model $b$ in having an external cooling 
flow in $r \gta 30$ kpc that is supported by gas 
heated by waves that reflect at this impedance barrier. 

The impedance barriers in models b and f, where the density and
temperature vary significantly over one wavelength, are a long term
result of unrealistically intense central wave-heating, and are
grossly incompatible with observations.  However, long before impedance
barriers appear in the solutions as shown in Figure 4, the temperature
and density structure of the hot gas has evolved far from typically
observed profiles.  In principle the hot central cores in these models
could be smoothed by outward thermal conduction, but this is irrelevant
since in observed clusters the inner temperature gradient is positive
and the conduction, if it were important, should be heating the
central gas, not cooling it.  The sharp positive density jump in
models b and f are very likely to be disrupted by Rayleigh-Taylor
instabilities.  However, in our previous 2D calculations, in which the
RT instability was allowed to develop naturally, we show that
including more computational dimensions does not save the problems
associated with the wave-heating hypothesis (Brighenti \& Mathews
2003). All centrally heated flows we considered (allowing for RT)
resulted in $dT/dr < 0$ in violation of {\bf CR2}.

\section{Summary and Conclusions}

Following the discovery of 
shock-ripples in the Perseus and Virgo clusters, 
it has often been assumed that this is the fundamental 
global heating mechanism that prevents the gas from cooling 
as required by X-ray spectral observations. 
A heating process can dynamically shut down a cooling flow 
if it distributes enough energy to reduce or 
reverse the small negative flow velocity out to approximately 
the cooling radius.
It is clear that the subsonic formation of large X-ray cavities 
creates weak shocks that move far out into the flow 
and that these waves dissipate some of their energy 
in remote regions of the flow.

Nevertheless, we have demonstrated here 
that the transport of energy by outwardly propagating waves 
is not likely to be the dominant process that accounts 
for the absence of cool and cooling gas in these flows.
The main problem with this idea is that the density structure
of the hot gas varies as $\rho \propto r^{-s}$ where $s < 2$ 
so that outgoing waves dissipate most of their energy in the 
central part of the flow and too little at $r \sim r_{cool}$.
The result is that after a few Gyrs the gas temperature 
flattens and eventually peaks near the center of 
the flow, just opposite to observed temperature profiles 
in cooling core clusters (e.g., Allen, Schmidt, \& Fabian 2001),
violating the second cardinal requirement ({\bf CR2}).
Much enthusiasm for ripple heating has resulted 
from calculations that proceeded for only $\sim0.3$ Gyrs
(e.g.  Ruszkowski, M., Br\"uggen, \& Begelman 2004a,b), 
as in our model $a1$,  but 
we have shown here that the temperature profile deteriorates 
on more relevant Gyr timescales.
In a similar calculation 
dalla Vecchia et al. (2004) describe wave-heated flows 
and compute to 1.5 - 5 Gyrs, but find, as we do, that 
the resulting temperature profile differs from those observed.

Very recently, after this paper was submitted, 
Fujita \& Suzuki (2005) presented a model of 
steady state cluster gas 
heating with weak shocks that reproduces many of the same 
results derived here.
In particular they find that a series of weak shocks 
propagating out in the Perseus Cluster results in 
unrealistically strong heating in the core.

The unsatisfactory results described here are similar 
to the difficulties we encountered in our earlier explorations  
of cooling flows that were heated in a variety of 
ways (Brighenti \& Mathews 2002a; 2003). 
Our centrally heated cooling flows produced rising bubbles 
and weak shocks that propagated to large radii, 
but we did not discuss distant wave dissipation in detail.
We also found that (feedback) 
heating the central regions of cooling flows 
often simply moved the cooling to larger 
radii but did not diminish the cooling rate 
(as in the ripple-heated models $b$ and $f$ described here).
Large scale cooling inflow can be stopped if the AGN 
feedback heating 
is rapid, extends to $\sim r_{cool}$ and is supplemented 
by thermal conduction from large radii 
(Ruszkowski \& Begelman 2002), but the amount that 
thermal conduction is suppressed by magnetic fields needs 
some fine-tuning (Brighenti \& Mathews 2003).
A similar unphysical fine-tuning is apparent in the 
wave-heated models described here -- the location, amplitude 
and period of wave creation by the X-ray cavities 
must be finely and unrealistically 
regulated to ensure just the proper amount 
of heating in distant regions of the flow 
to balance radiative losses,
i.e. $E_m \sim E_{rad}$,
although the resulting temperature gradient is unacceptable. 

Most of this fine-tuning can be avoided if both mass 
and energy are redistributed outward in the hot gas as
in the ``circulation flows'' described by 
Mathews et al. (2003) and Mathews, Brighenti \& Buote (2004).
In these flows the mass transported by dense inflowing gas 
is balanced by a nearly equal amount of gas flowing outward 
to large radii carrying both mass and energy.
The denser inflow emits most of the radiation and appears 
as a normal cooling flow with a temperature profile that 
matches the observations.
Meanwhile, bubbles of hot gas heated by the central AGN flow out 
rapidly and contribute much less to the overall 
X-ray emission. 
Our previous 2D computational models of heated flows lacked the 
spatial resolution to isolate individually rising 
bubbles as they move large distances upstream 
through the denser gas, 
but circulation flows can be computed in a 
more schematic fashion (Mathews, Brighenti \& Buote 2004).
In mass circulating flows it is only necessary that gas near 
the central AGN be heated sufficiently to have 
an entropy similar to gas at approximately $r_{cool}$ 
or beyond. 
The best argument for outward mass circulation are the 
vast regions of iron-enriched gas that extend far 
beyond the stars in the central galaxy where the 
iron is produced.

We argue here that neither the $PV$ work done as 
the hot X-ray cavities inflate or viscous heating 
by cavity-driven turbulence is a strong source of local heating;
this is consistent with the low gas temperatures 
observed in cooling-core clusters in the central region 
where bubbles are observed. 
This suggests that the AGN energy (and mass) 
must be distributed non-locally to more  
distant gas.

To summarize, we have shown that viscous dissipation 
in outward propagating weak shock waves cannot 
keep the gas in the Perseus 
cluster from cooling without heating the gas in the 
central regions far above the observed temperatures.
Since weak shocks have been observed in Perseus and 
elsewhere, some heating is expected but they are not 
the dominant heating mechanism.
The shocks observed are almost certainly produced 
by the subsonic inflation of X-ray cavities, 
which, as we have shown, produce non-linear waves 
that rapidly steepen into weak shocks.
Evidently cavity-produced shocks are too infrequent or weak 
to generate mechanical luminosities comparable to 
the X-ray losses.

\vskip.1in

Studies of the evolution of hot gas in elliptical galaxies
at UC Santa Cruz are supported by
NASA grants NAG 5-8409 \& ATP02-0122-0079 and NSF grants
AST-9802994 \& AST-0098351 for which we are very grateful.

\clearpage

\makeatletter
\def\jnl@aj{AJ}
\ifx\revtex@jnl\jnl@aj\let\tablebreak=\nl\fi
\makeatother

\begin{deluxetable}{rcccccc}
\tablewidth{0pc}
\tablecaption{PISTON PARAMETERS TO CREATE SHOCK-RIPPLES}
\tablehead{
\colhead{model} &
\colhead{$t_{comp}\tablenotemark{a}$} &
\colhead{$r_p$} &
\colhead{$\Delta r_p$} &
\colhead{$T_p$} &
\colhead{$|u_p|$} &
\colhead{$r_p^2|u_p|$} \\
\colhead{} &
\colhead{(Gyr)} &
\colhead{(kpc)} &
\colhead{(kpc)} &
\colhead{(yr)} &
\colhead{(km/s)} &
\colhead{(kpc$^2$ km/s)}
}
\startdata
a1  &  0.04  & 23.8  & 3.9  & $3 \times 10^7$ & 800 & $4.5 \times 10^5$
\cr
a  &  4  & 23.8  & 3.9  & $3 \times 10^7$ & 800 & $4.5 \times 10^5$
\cr
b  &  4  & 23.8  & 1.3  & $1 \times 10^7$ & 800 & $4.5 \times 10^5$
\cr
c  &  4  & 31.5  & 2.23  & $3 \times 10^7$ & 457 & $4.5 \times 10^5$
\cr
d  &  4  & 23.8  & 2.23  & $3 \times 10^7$ & 457 & $2.6 \times 10^5$
\cr
e  &  4  & 18.1  & 2.23  & $3 \times 10^7$ & 457 & $1.5 \times 10^5$
\cr
f  &  4  & 13.8  & 2.23  & $3 \times 10^7$ & 457 & $0.87 \times 10^5$
\cr
\enddata
\tablenotetext{a}{Total time of flow calculation.}
\end{deluxetable}


\clearpage
\begin{figure}
\centering
\includegraphics[bb=90 216 522 569,scale=0.9,angle= 270]
{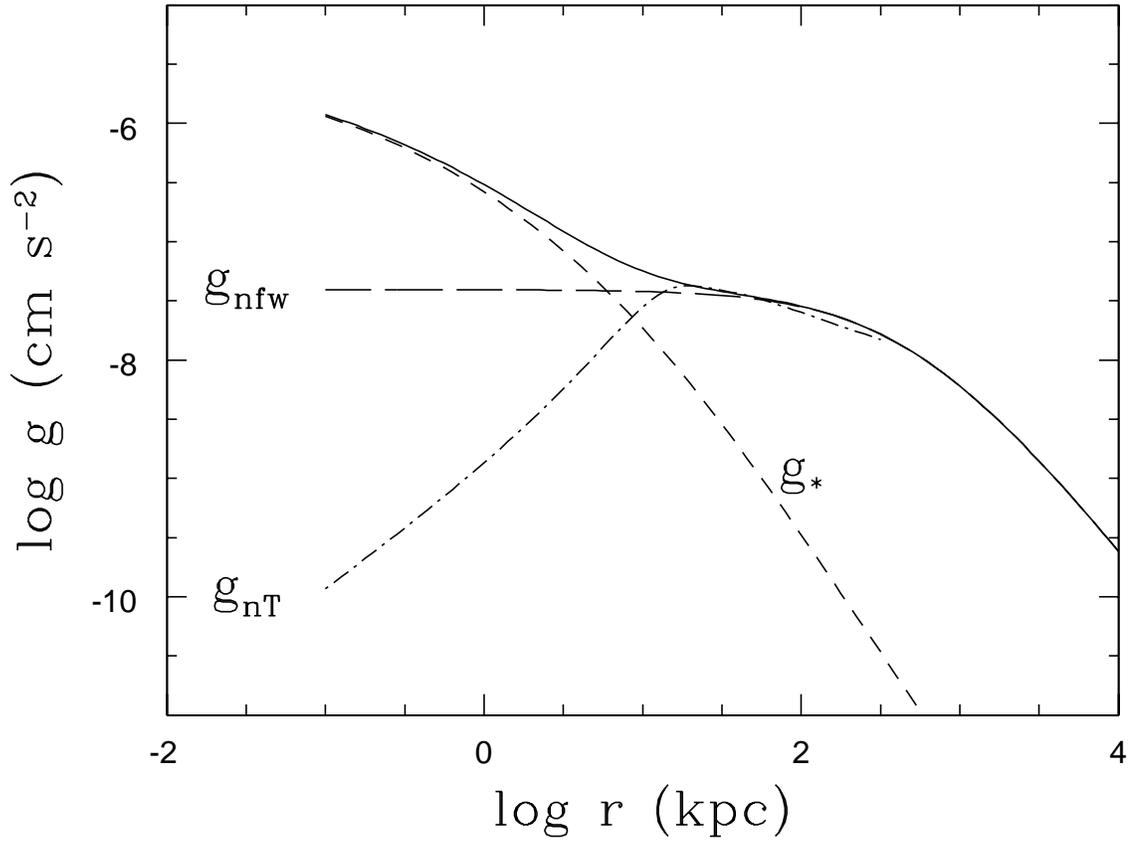}
\vskip.7in
\caption{
Acceleration of gravity in the Perseus cluster: 
$g_{nT}$ derived from the observed gas density and temperature 
(dot-dashed line), $g_{nfw}$ component from the dark halo 
(long dashed line),
$g_*$ component from stars in the central galaxy NGC 1275
(short dashed line) and 
$g = g_{nfw} + g_*$ is the total adopted acceleration 
(solid line).
}
\label{fig1}
\end{figure}

\clearpage
\begin{figure}
\vskip2.in
\includegraphics[bb=90 216 522 569,angle= 270]
{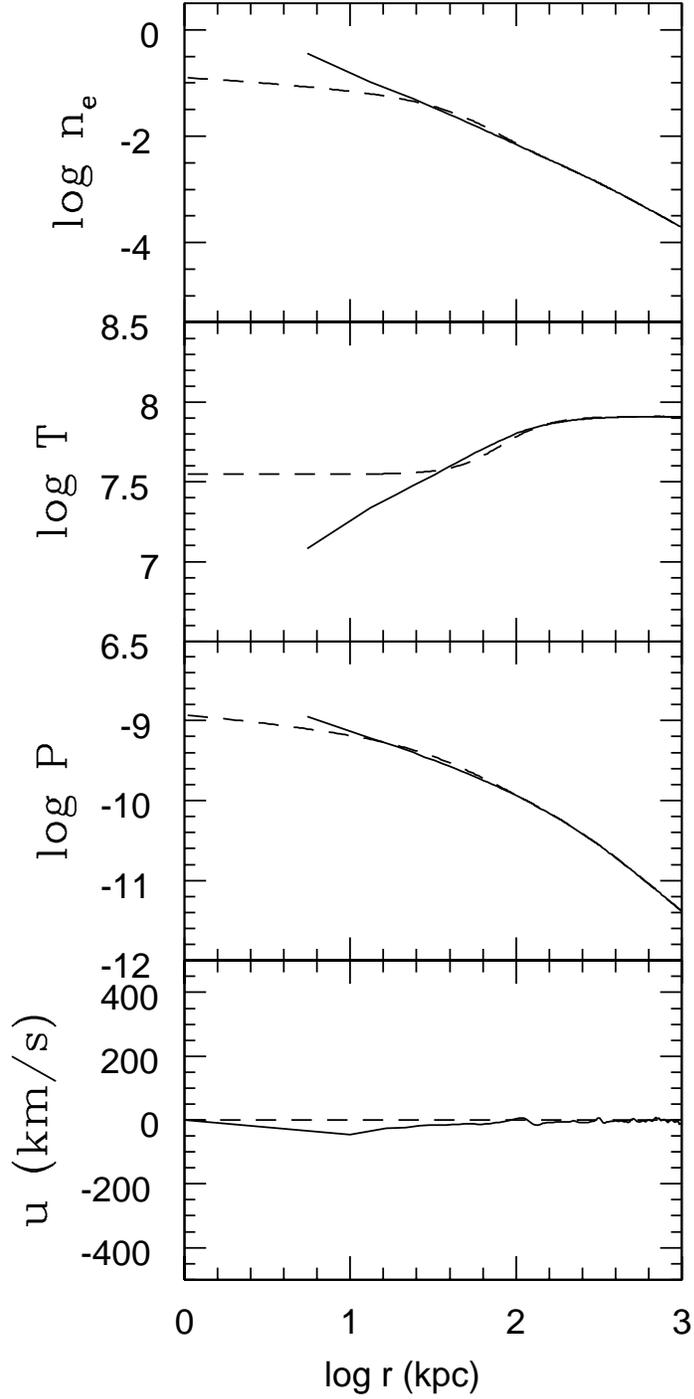}
\vskip.7in
\caption{
A standard cooling flow in the Perseus cluster after 
4 Gyrs (solid lines) 
overplotted with the observed gas density, 
temperature and pressure (dashed lines) and initial 
velocity (dashed line).
}
\label{fig2}
\end{figure}

\clearpage
\begin{figure}
\centering
\includegraphics[bb=90 216 522 569,scale=0.9,angle= 270]
{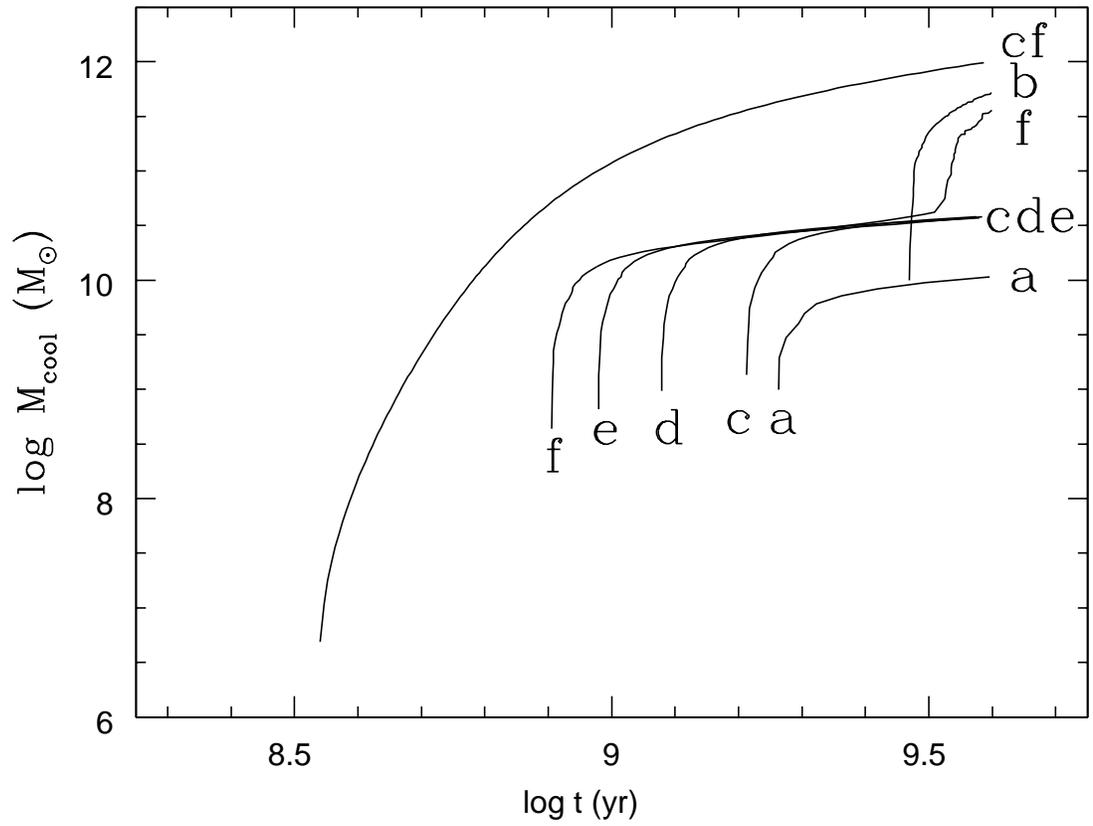}
\vskip.7in
\caption{
Total mass of gas in $M_{\odot}$ 
that cooled after time $t$ in the 
cooling flow (model $cf$) and in six flows with ripple-shock 
heating, models $a \rightarrow f$.
}
\label{fig3}
\end{figure}

\clearpage
\begin{figure}
\vskip2.5in
\includegraphics[scale=0.7,angle= 270]
{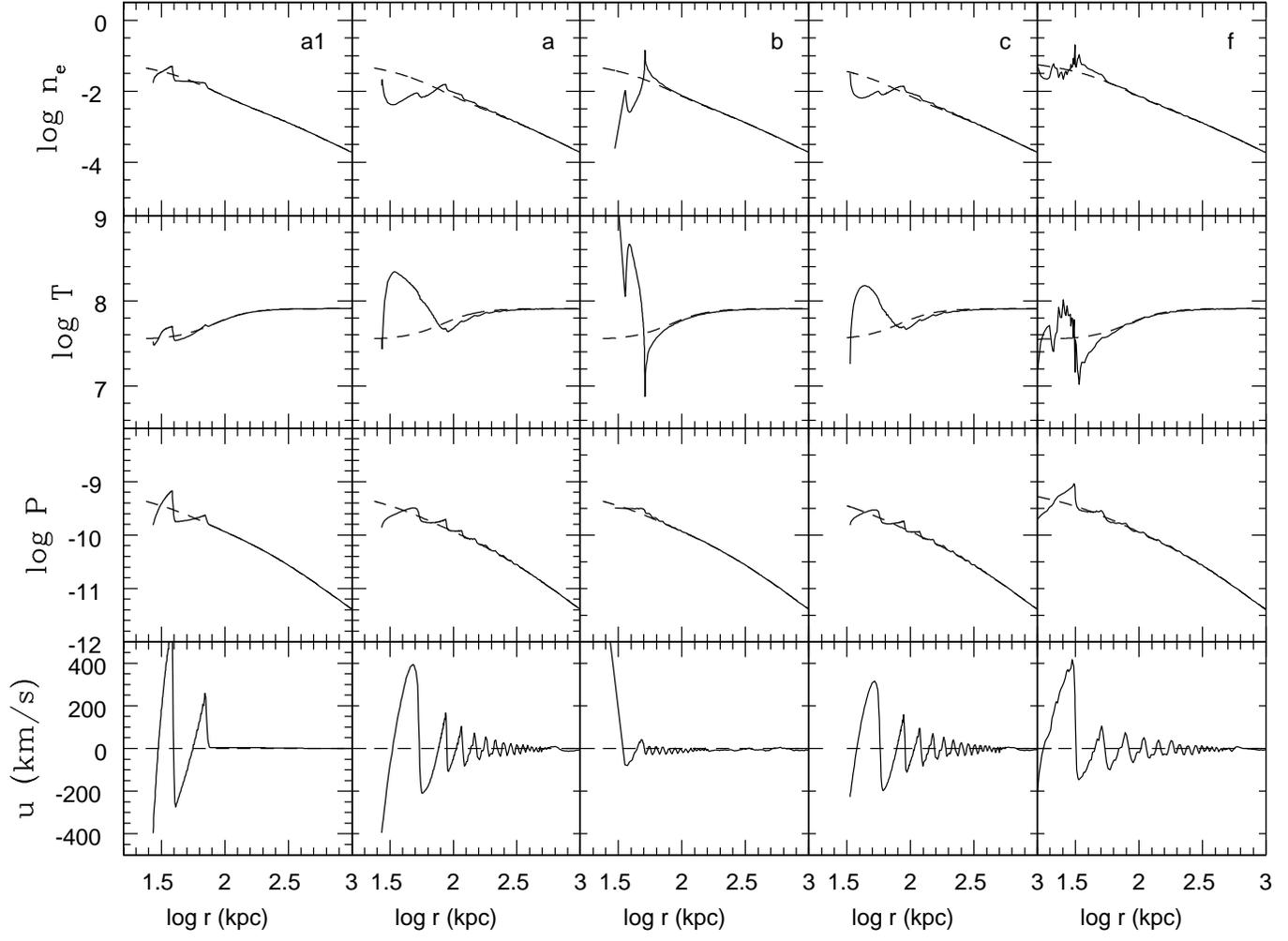}
\vskip.7in
\caption{
Computed gas density, temperature, pressure and 
velocity after 4 Gyrs in models $a - c$ and $f$ 
(solid lines) compared with the observed (or initial) 
profiles (dashed lines).  
}
\label{fig4}
\end{figure}

\clearpage
\begin{figure}
\vskip2.in
\includegraphics[bb=90 216 522 569,angle= 270]
{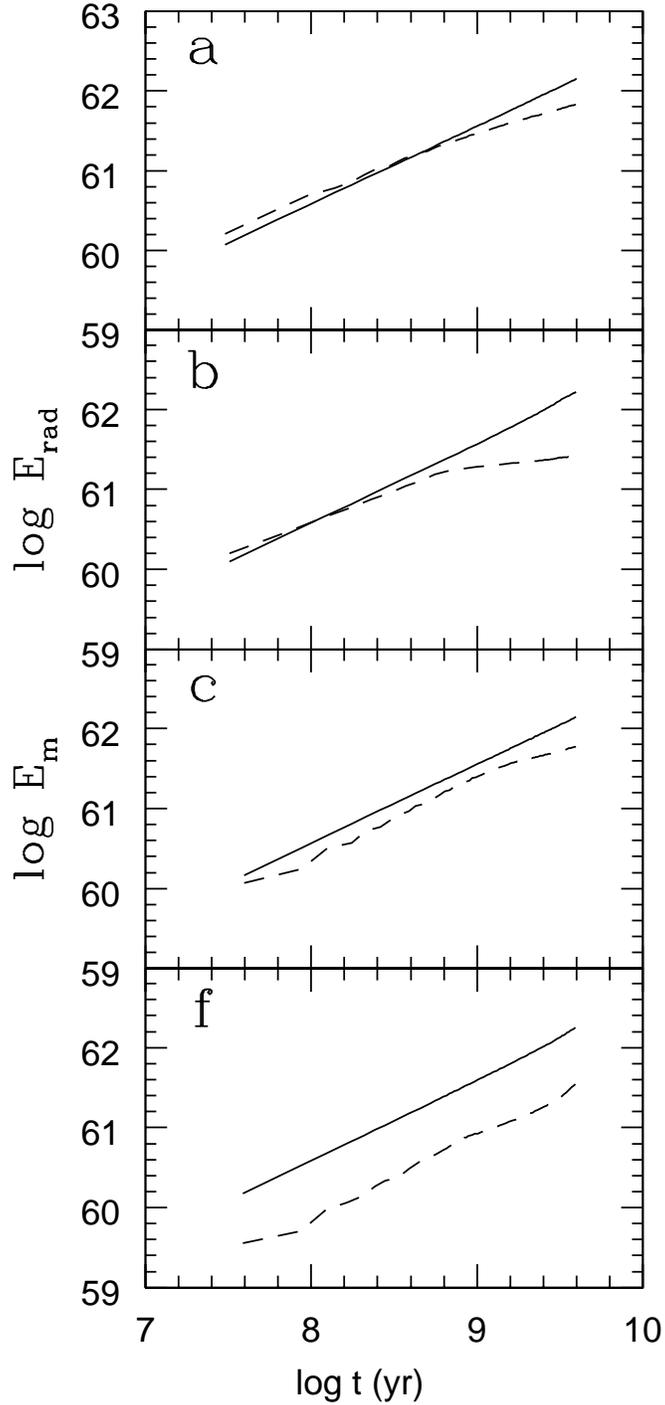}
\vskip.7in
\caption{
Total amount of mechanical 
energy $E_m(t)$ delivered to the Perseus gas 
by the spherical piston after 
time t (dashed lines) for flow models 
$a - c$ and $f$ is compared to the total bolometric 
energy lost in radiation $E_{rad}(t)$ 
(solid lines), both in ergs.
}
\label{fig5}
\end{figure}

\end{document}